# Cavitation in dielectric fluid in inhomogeneous pulsed electric field

**M. N. Shneider[1,*] and M. Pekker[2]**

[1] *Department of Mechanical and Aerospace Engineering, Princeton University, Princeton, NJ 08544, USA*

[2] *Drexel Plasma Institute, Drexel University, 200 Federal Street, Camden, NJ 08103, USA*

## Abstract

This paper proposes a method for studying the early stages of the cavitation development in arbitrary, non-stationary conditions. This method is based on the comparison of the results of calculations in the framework of a theoretical model of the liquid dielectrics motion in a strong non-uniform electric field and experiments with controlled parameters. This approach allows us to find the critical negative pressure, at which cavitation begins to develop, and to determine the values of the constants in the classical models of cavitation.

## Introduction

The cavitation is a hydrodynamic phenomenon, which has not lost its currency for more than a century of research. The problem of the collapse of the bubble in liquid was first considered by Rayleigh in the early 20th century [1]. However, this theory does not answer the question about the cause of these bubbles. A theory of bubbles (cavitation voids) formation due to thermal fluctuations was developed by Zel'dovich [2] and Fisher [3]. Later, the formula obtained in [3] for the negative pressure, at which cavitation voids are formed (see below), was not modified, because, as shown in [3], the result is weakly dependent on the waiting time for the "bubble" appearance and the observation volume. The main efforts of researchers have been spent on the experimental study of the conditions for formation of cavitation voids. Numerous experiments had shown that the negative pressure, at which cavitation voids are formed, is significantly different from the theoretically predicted, and is strongly dependent on the fluid temperature, presence of impurities, the amount of dissolved gas, and many other parameters (see, e.g., [4,5]). It should also be noted that the various methods of measuring the critical pressure, at which cavitation occurs, give very inconsistent results. For example, cavitation in water was observed in [6] when the absolute value of negative pressure was less than 1 MPa. While in [4, 7-12] the observed threshold of cavitation ranged from -1 MPa to -40 MPa. We believe that this is due to the fact that the region, in which cavitation does occur, is rather uncertain, and the critical size of the bubbles being small (of the order of nanometers) cannot be detected. Only the developed bubbles with sizes of the order of a micron are easily observed in the experiments.

A mechanism for the rapid breakdown in the fluid, associated with the occurrence of cavitation ruptures under the influence of electrostrictive forces near the needle electrode, was proposed in [13]. Later, the hydrodynamic model of compressible fluid motion under the influence of the ponderomotive electrostrictive forces in a non-uniform time-dependent electric field was suggested in [14]. As shown in [14], if the voltage on the needle electrode is growing fast enough (a few nano seconds), the negative stress is created in liquid, and it may be sufficient for the cavitation formation. A nanosecond breakdown, the beginning of which can be explained by the formation of cavitation in a stretched liquid due to electrostriction, was investigated experimentally in [15] - [19]. Later stages of the evolution of gas-filled bubbles in a dielectric liquid within a strong electric field are theoretically and experimentally studied in [20], [21].

* m.n.shneider@gmail.com

In this paper, we propose a method that allows to determine the critical parameters, at which cavitation begins, on the basis of a comparison between the experiment and the simulation results within the framework of hydrodynamics of compressible fluids.

## 1. Probability of cavitation void formation in the absence of electric field

The energy required to create a bubble of the radius $R$ in the liquid is equal:

$$\Sigma(R) = \frac{4}{3}\pi R^3 \left(P - P_{sat}\right) + 4\pi R^2 \sigma \, . \tag{1}$$

Here, $P_{sat}$ is the vapor pressure in the bubble, $P$ is the external pressure, $\sigma$ is the surface tension of the liquid. The pressure $P$ is negative in the case of tensile stress. From (1) it is easy to find the critical size of the bubble, after which it will expand, and the work that must be done for its creation:

$$R_{cr} = \frac{2\sigma}{P_{sat} - P}, \quad \Sigma_{cr} = \frac{16\pi\sigma^3}{3\left(P_{sat} - P\right)^2} \, . \tag{2}$$

Further, we neglected $P_{sat}$ in the formula (2), since the absolute value of the negative pressure, at which the cavitation bubbles form, is much higher than the saturated vapor pressure. The probability of appearance of the critical bubble in the volume $V$ over the time $t$ due to development of thermal fluctuations, in accordance with the theory [2,3], is

$$W_{pore} = 1 - \exp\left( -\int_0^t \int_V \Gamma dt_1 d\vec{r} \right) = 1 - \exp\left( -\int_0^t \int_V \frac{1}{V_{cr} \cdot \Delta t} \exp\left( -\frac{\Sigma_{cr}}{k_B T} \right) dt_1 d\vec{r} \right) . \tag{3}$$

Here, $\Gamma$ [$m^{-3}s^{-1}$] characterizes the rate of the cavitation voids appearance in unit volume per second. The integration in (3) is carried out in time and over the entire area of observation. Here, $T$ is the temperature of fluid in Kelvin, $k_B$ is the Boltzmann constant, $V_{cr} = \frac{4}{3}\pi R_{cr}^3$ is the volume of the critical bubble. The time interval $\Delta t$ is determined from the Heisenberg uncertainty principle: $\Delta t \cdot \bar{\varepsilon} = 2\pi\hbar$, where $\bar{\varepsilon}$ is the mean thermal energy of a water molecule. After all the corresponding substitutions, the value $\Gamma$ in formula (3) has the form:

$$\Gamma = \frac{3k_B T}{16\pi(\sigma \cdot k_\sigma)^3} \frac{|P|^3}{4\pi\hbar} \exp\left( -\frac{16\pi(\sigma \cdot k_\sigma)^3}{3k_B T \cdot P^2} \right). \quad (4)$$

In the formula (4), in accordance with [20], the correction factor to the surface tension of liquid is introduced:

$$k_\sigma = \frac{1}{1 + 2\delta / R_{cr}}, \quad (5)$$

where $\delta = \rho_S / (\rho_{fluid} - \rho_{vapor})$ is the so-called Tolman coefficient [22]. $\rho_S$ is the superficial density of fluid at the boundary, $\rho_{fluid}$ and $\rho_{vapor}$ are the volume densities of the fluid in the interior of the liquid and vapor phases. The coefficient $\delta$ is determined experimentally. Note that in the recent paper [12], on the basis of statistical analysis of the experimental data, the correctional formula for the surface tension as a function of the radius of the cavitation bubble has been proposed, and, thus, the empirical Tolman coefficient $\delta$ was determined.

## 2. Cavitation void formation in strong electric field

In a strong non-uniform electric field, the negative pressure in a dielectric liquid created by the electrostrictive forces [23], [24] is given by [25], [14]:

$$P = -\frac{1}{2}\left( \frac{\partial \varepsilon}{\partial \rho} \rho \right) \varepsilon_0 E^2 = -\frac{1}{2} \alpha \varepsilon \varepsilon_0 E^2 \quad (6)$$

Here, $\varepsilon_0$ is the vacuum permittivity, $\varepsilon$ is the dielectric constant of the liquid, $\alpha$ is the constant, independent of the electric field. For non-polar dielectric liquids: $\alpha = (\varepsilon - 1) \cdot (\varepsilon + 2) / 3$ [23], and for most of the studied polar liquids, including water, $\alpha \approx 1.3 - 1.5$ [25], [26]. However, the formation of cavitation void results in changing electric field in the void vicinity as well as inside the void. As a consequence, the expression for the negative pressure should be recalculated taking into account the discontinuity of the dielectric constant at the liquid-vacuum boundary. Since the tangential component of the electric field and the normal component of induction ($\vec{D} = \varepsilon \varepsilon_0 \vec{E}$ )are continuous at the boundary of the bubble, the expression for the surface forces exerted by the electric field on the surface of the pore per unit area is [24]:

$$F_{S,n} = \frac{\varepsilon_0}{2} \left( \alpha \left( \frac{E_{p,n}^2}{\varepsilon} + \varepsilon E_{p,t}^2 \right) - (\varepsilon - 1) \left( E_{p,t}^2 + \frac{E_{p,n}^2}{\varepsilon} \right) \right), \quad [\mathrm{N/m^2}]. \quad (7)$$

Here $E_{p,n}$, $E_{p,t}$ are the normal and tangential components of the electric field inside the pore. Since the critical initial sizes of the cavitation pores are few nanometers, and the field is changing on a length scale of 1-10 microns, a field in the vicinity of the void can be considered as a homogeneous with good accuracy. In this case, the electric field is uniform inside the ellipsoidal pore [27]. Wherein the expressions for the normal and tangential components of the electric field on the surface of the spherical pore are [24], [27]:

$$E_{p,n} = \frac{3\varepsilon}{1 + 2\varepsilon} E \cos(\theta), \quad E_{p,t} = \frac{3\varepsilon}{1 + 2\varepsilon} E \sin(\theta) \quad (8)$$

Where $E$ is undistorted electric field; if the void is absent, $\theta$ is the azimuthal angles in a spherical coordinate system. The total force per unit area acting on the surface of the pores is formed by (7) and the surface tension force, $F_n = F_{S,n} - 2\sigma / R$. By substituting (7) into (6) we find the energy needed to create the pore of the radius $R$:

$$\Sigma(R) = 2\pi \int_0^R \int_0^\pi r^2 F_n \sin(\theta) d\theta dr = 4\pi R^2 \sigma - \left(\frac{3\varepsilon}{1+2\varepsilon}\right)^2 \frac{\pi R^3}{3} \varepsilon_0 E^2 \left(\frac{4(\alpha-1)\varepsilon}{3} + \frac{2(\alpha+1)}{3\varepsilon} + \frac{2}{3}\right). \quad (9)$$

Since for water $\varepsilon = 81 >> 1$, the expression (9) with a good accuracy can be converted to the form:

$$\Sigma(R) = 4\pi R^2 \sigma - \pi R^3 (\alpha - 1)\varepsilon_0 \varepsilon E^2 = 4\pi R^2 \sigma + 2\pi R^3 \widetilde{P}, \quad (10)$$

where $\widetilde{P} = -\frac{1}{2}(\alpha - 1)\varepsilon_0 \varepsilon E^2$ is an effective negative electrostriction pressure, stretching the pore.

From (10) we obtain the critical radius and the energy required to create the cavitation pore:

$$R_{cr} = \frac{4}{3}\frac{\sigma}{\left|\widetilde{P}\right|}, \quad \Sigma_{cr} = \pi \frac{64}{27}\frac{\sigma^3}{\widetilde{P}^2} \quad (11)$$

It should be noted that this consideration is valid in the case when the concentration of voids is relatively low: when average distance between the pores is much greater than their size.

## 3. Probability of cavitation void formation in strong electric field

Let us calculate the probability of a void creation in strong inhomogeneous electric field for the simplest case of a spherical electrode with a growing linearly voltage pulse applied. Using $R_{cr}, \Sigma_{cr}$ (11) one can obtain the expression for $\Gamma$, similar to (4):

$$\Gamma = \frac{3\left|\widetilde{P}\right|}{8\pi\hbar} \cdot \frac{27}{64\pi} \frac{\left|\widetilde{P}\right|^2 k_B T}{(\sigma k_\sigma)^3} \exp\left(-\frac{64\pi}{27}\frac{(\sigma k_\sigma)^3}{\left|\widetilde{P}\right|^2 k_B T}\right). \quad (12)$$

We will consider a relatively short pulse with a steep front, such that the fluid does not have time to come in motion (we will consider below how steep the pulse should be). The electric field near the electrode varies as:

$$E = \frac{U_0 r_0}{r^2}\frac{t}{t_0}, \quad (13)$$

where $U_0$ is the amplitude of the voltage, $r_0$ is the radius of the electrode, $t_0$ is the duration of the pulse front. The field (13) in the dielectric fluid due to the electrostrictive effect, [23], [24] leads to the appearance of negative pressure region [25, 14]. Thus, the negative pressure stretching the microspores, which are forming as a result of the thermal fluctuations is given by:

$$\widetilde{P} = -\frac{1}{2}(\alpha-1)\varepsilon\varepsilon_0 E^2 = -\frac{1}{2}(\alpha-1)\varepsilon\varepsilon_0 \frac{(U_0 r_0)^2}{r^4}\frac{t^2}{t_0^2}, \qquad (14)$$

Substituting (14) into (4) we get:

$$\psi = \ln(1-W_{pore}) = -\frac{2C}{B_0 k_\sigma^3} P_0^3 r_0^3 t_0 \int_0^{\tau=1} d\tau \int_1^{\eta_1} \frac{\tau^6}{\eta^{10}} \exp\left(-\frac{B_0 k_\sigma^3}{P_0^2}\frac{\eta^8}{\tau^4}\right) d\eta . \qquad (15)$$

Here:

$$B_0 = 10^{-12}\frac{64\pi\sigma^3}{27 k_B T},\ C = \frac{3\cdot 10^6}{4\hbar},\ \widetilde{P}_0 = 10^{-6}\left((\alpha-1)\frac{\varepsilon_0\varepsilon}{2}\right)\left(\frac{U_0}{r_0}\right)^2,$$
$$0 \le \tau = \frac{t}{t_0} \le 1,\quad \eta_1 = \frac{r}{r_0} \ge 1,\quad \alpha = 1.5 \qquad (16)$$

The numeric constants are selected in such a way that the "electrostrictive" pressure $P_0$ in (15) and (16) is measured in MPa.

Since $\psi$ in (15) is nothing else but the logarithm of the probability that during the time $t$ the cavitation bubbles of critical size do not appear in the volume $V = \frac{4}{3}\pi(r^3 - r_0^3)$, that at $\eta_1 = \frac{r}{r_0} \to \infty$ and $|\psi| < 1$ - the probability of formation of cavitation bubbles is small, but if $|\psi| > 1$ - the probability of formation of cavitation bubbles is great. It is convenient to use the variables:

$$\theta = \frac{P_0^{1/2}}{B_0^{1/4} k_\sigma^3 \eta^2},\quad y = \theta\tau . \qquad (17)$$

In these variables, the expression (8) takes the form:

$$\psi = -C \cdot P_0 r_0^3 t_0 \cdot \frac{1}{\sqrt{\xi}} \int_0^{\xi} \frac{1}{\theta^{7/2}} d\theta \int_0^{\theta} y^6 \exp\left(-\frac{1}{y^4}\right) dy , \qquad (18)$$

where $\xi = \frac{P_0^{1/2}}{B_0^{1/4} k_\sigma^{3/4}}$ .

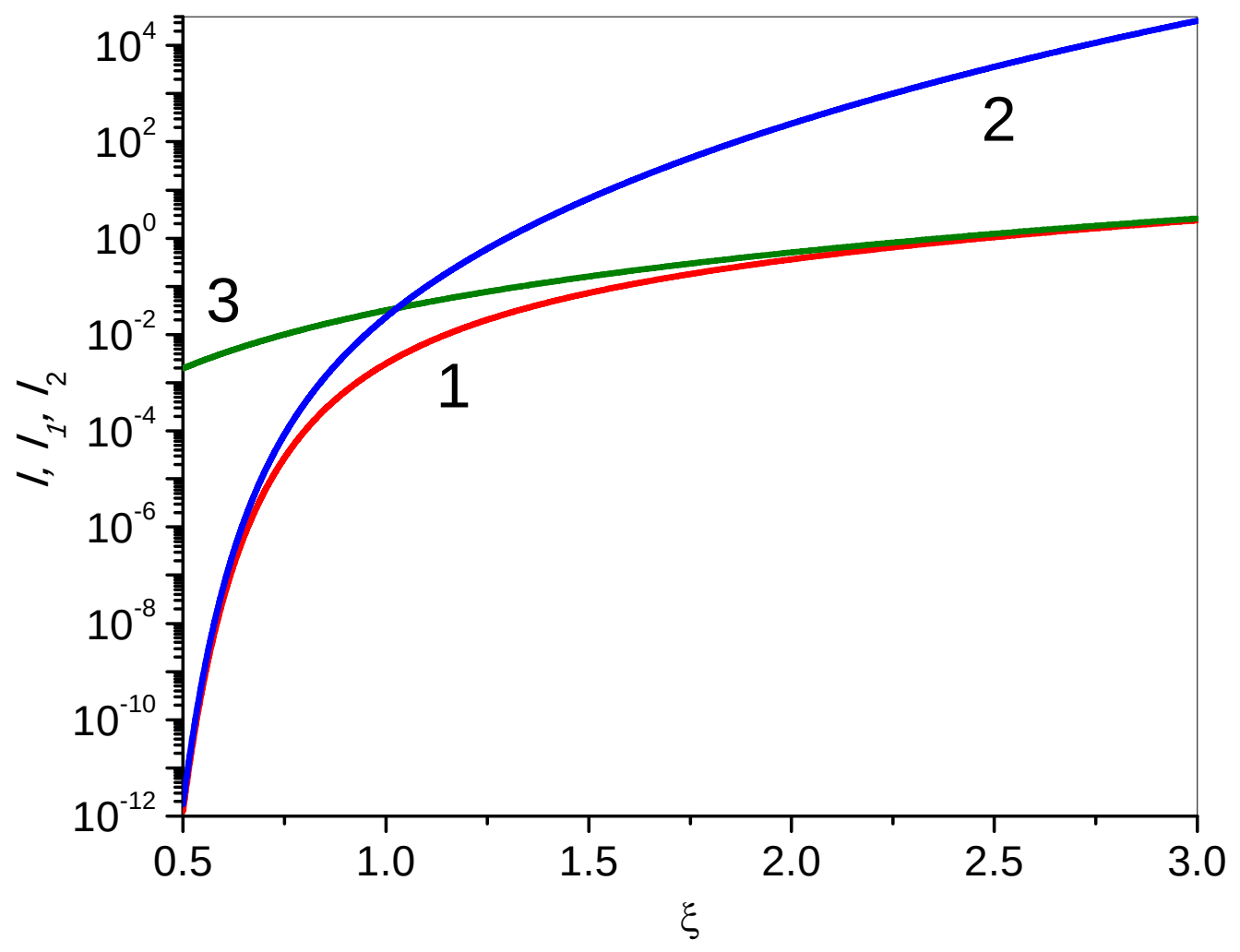

**Figure 1.** The dependence of the integral $I$ on the upper limit of integration $\xi$ (curve 1); 2 and 3 are the asymptotes $I_1$ and $I_2$, correspondingly.

The integral in (11) $I = \frac{1}{\sqrt{\xi}} \int_0^{\xi} \frac{1}{\theta^{7/2}} d\theta \int_0^{\theta} y^6 \exp\left(-\frac{1}{y^4}\right) dy$ has the asymptotes

$$I_1 = \frac{2}{63}\xi^4 \text{, at } \xi > 1.5 \tag{19}$$

and

$$I_2 = \frac{1}{16}\xi^{12} \exp\left(-\frac{1}{\xi^4}\right) \text{, at } \xi < 0.75 \; . \tag{20}$$

The dependences $I$ , $I_1$, $I_2$ on $\xi$ are shown in Figure 1.

Expression (12) is, in fact, the equation for the "electrostrictive" pressure $P_0$, at which the condition $|\psi| = 1$ is fulfilled.

Figure 2 shows the values of $P_0$ and the corresponding values of the voltage amplitude at the electrode $U_0$, at which the cavitation occurs at the assumed parameters $t_0 = 3$ ns, $r_0 = 50$ μm and $r = \infty$. The surface tension of water at $T = 293$ K is $\sigma = 0.072$ [N/m] [28]. The dependence of parameter $\xi$ , at which in the equation (18) holds $|\Psi| = 1$ is shown in Fig. 3. The corresponding dependence of the critical radius $R_{cr}$ on $k_\sigma$ is shown in Fig. 4. It is seen that the value $\xi$ , at which $|\Psi| = 1$ is almost independent of $k_\sigma$ , but the critical radius reduces in three times when $k_\sigma$ changes in the range from 0.1 to 1. If we can experimentally determine the value $P_0$ , then we can determine the correction $k_\sigma$ to the surface tension coefficient $\sigma$.

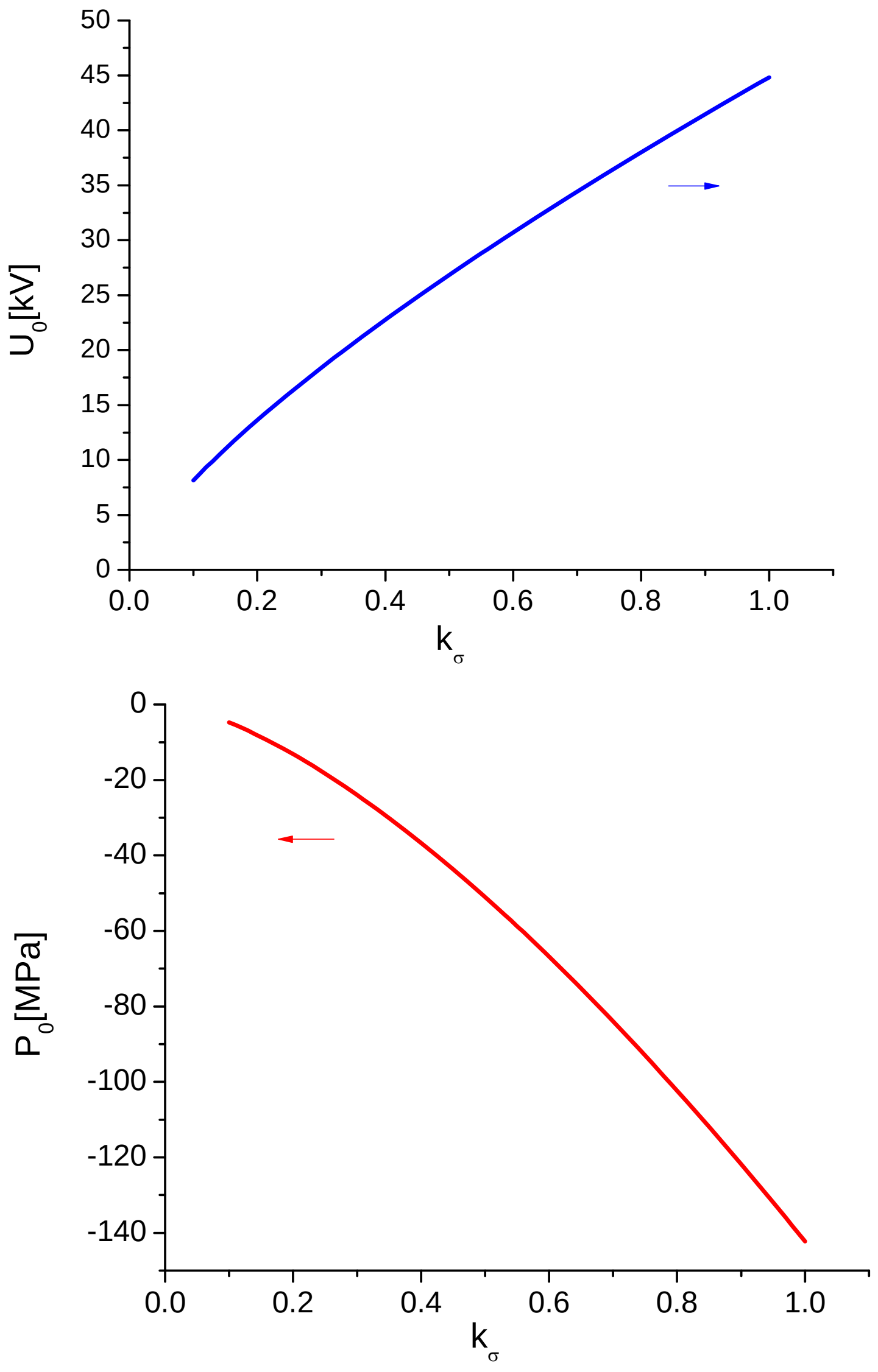

**Figure 2.** The dependence on $k_\sigma$ of the "electrostrictive" negative pressure $P_0$ and the corresponding amplitude of the voltage on a spherical electrode $U_0$, at which cavitation occurs at the assumed parameters $t_0 = 3$ ns, $r_0 = 50$ μm.

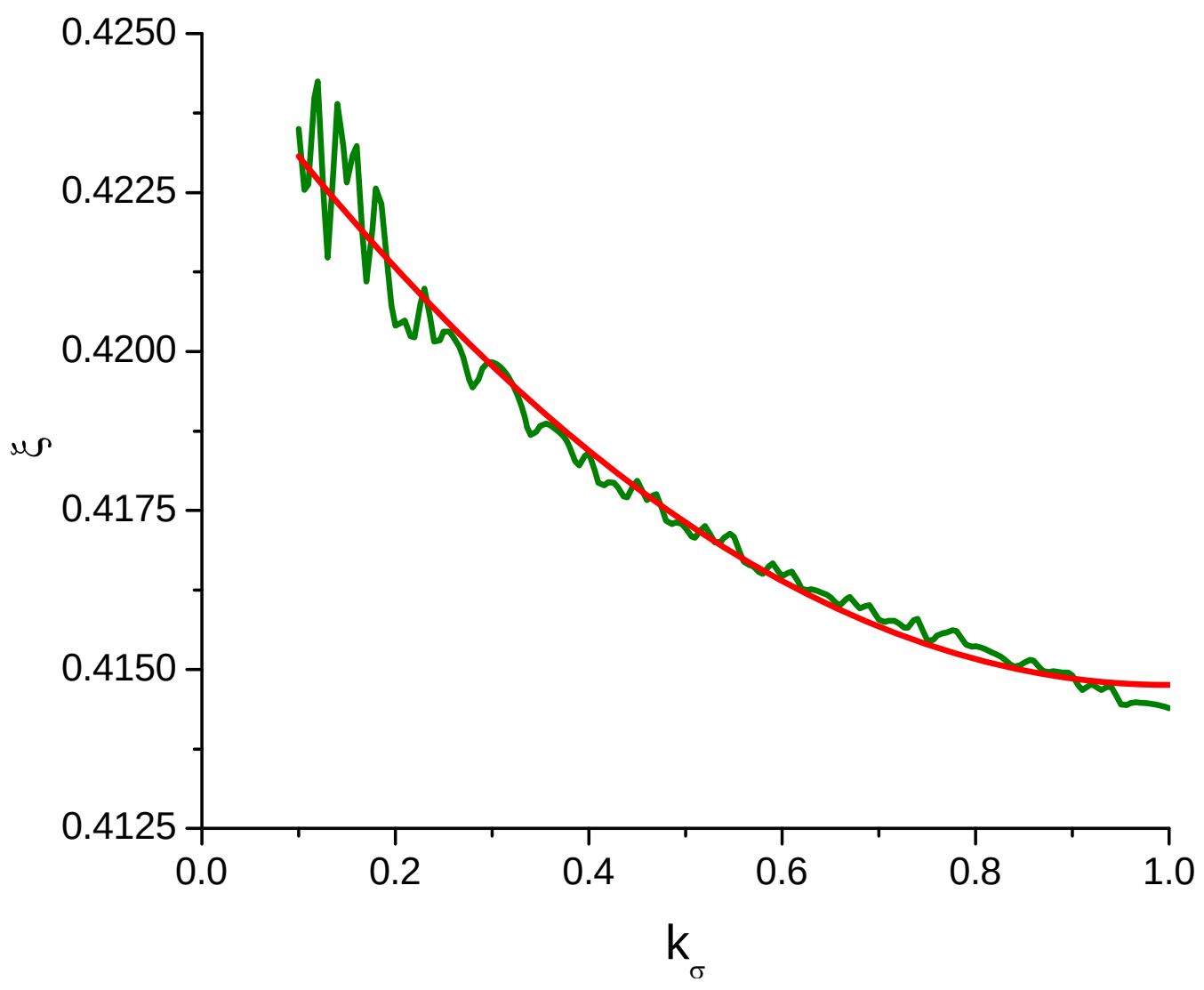

**Figure 3.** The dependence of the parameter $\xi$ , at which the condition for the cavitation beginning, $|\Psi| = 1$, is valid.

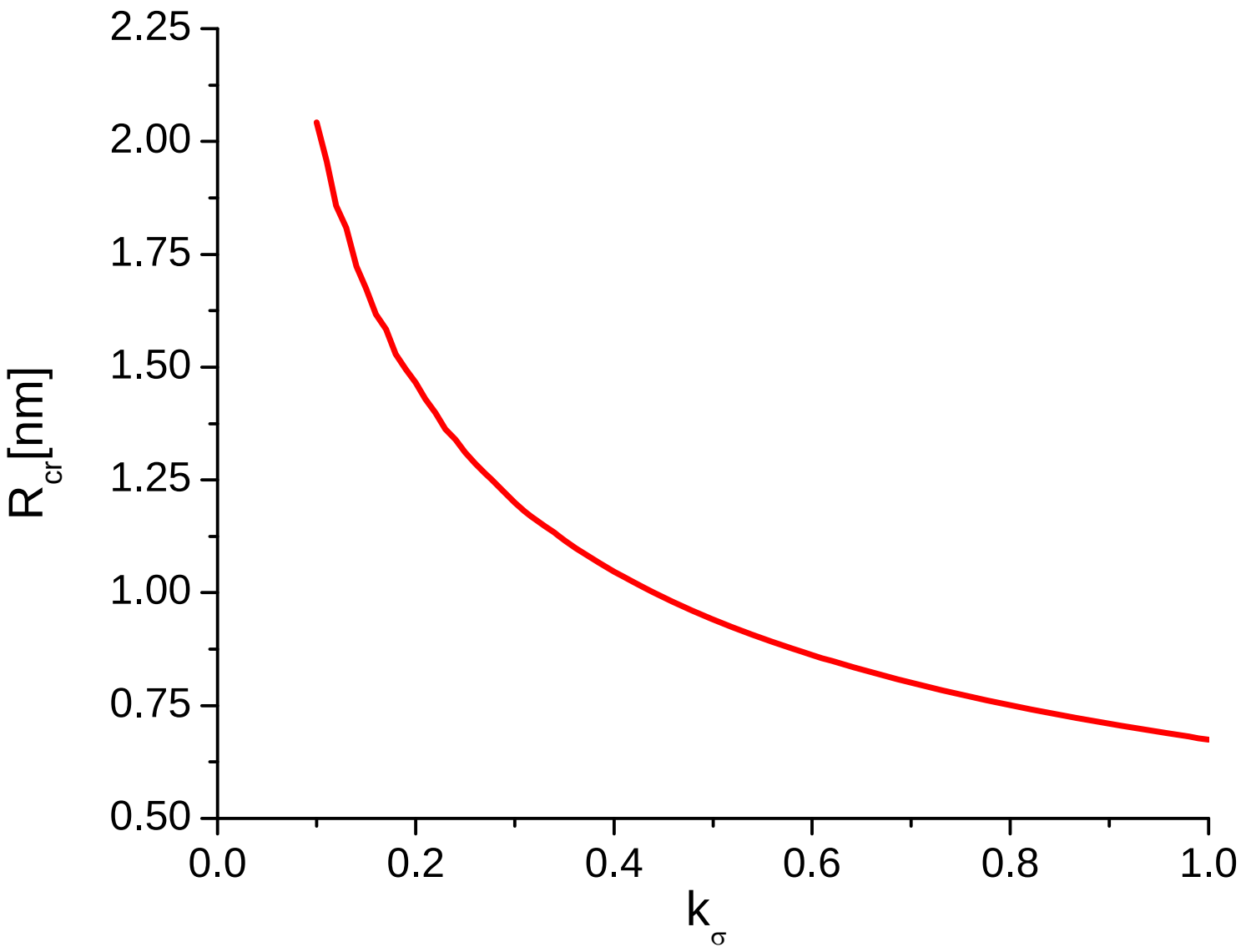

**Figure 4.** The dependence of the critical radius $R_{cr}$ on $k_\sigma$ .

The above calculations are valid when the electrostrictive ponderomotive forces acting on a dielectric fluid do not have enough time to put it in motion, that is, the voltage pulse front is short enough. The conditions when the motion of the fluid can be neglected are discussed below.

**4. Motion of the fluid under the action of the ponderomotive forces**.

The equations describing the dynamics of a liquid dielectric (in our case, water) in a pulsed inhomogeneous electric field in the approximation of a compressible fluid [14, 29] for the spherically symmetric case:

$$\frac{\partial u}{\partial t} = -\frac{1}{2}\frac{\partial}{\partial r}u^2 - \frac{1}{\rho}\frac{\partial}{\partial r}\left(p - \frac{\alpha}{2}\varepsilon_0 \varepsilon E^2\right) + \nu\left(\frac{4}{3}\frac{\partial^2 u}{\partial r^2} + \frac{4}{3}\frac{1}{r}\frac{\partial u}{\partial r} - \frac{2}{3}\frac{u}{r^2}\right)$$

$$\frac{\partial \rho}{\partial t} = -\frac{1}{r^2}\frac{\partial}{\partial r}\left(r^2 \rho u\right) \tag{21}$$

and the Tait equation of the state, which relates the pressure to the density of water [30],[31]:

$$p = \left(p_0 + B\right)\left(\frac{\rho}{\rho_0}\right)^{\gamma} - B \ , \tag{22}$$

$$\rho_0 = 1000 \ \text{kg/m}^3, \quad p_0 = 10^5 \ \text{Pa}, \quad B = 3.07 \cdot 10^8 \ \text{Pa}, \quad \gamma = 7.5$$

Here, $\rho$ is the fluid density, $p$ is the pressure, $u$ is the velocity, $\nu = 10^{-6}$ $m^2$/s is the kinematic viscosity of water.

The results of calculations for the different values of the rise time of the voltage pulse $t_0$ = 3, 10, and 40 ns. The calculations are made for the following parameters: the radius of the electrode $r_0 = 50\ \mu\text{m}$, and the voltage amplitude ,$U_0 = 22.4\ \text{kV}$ .

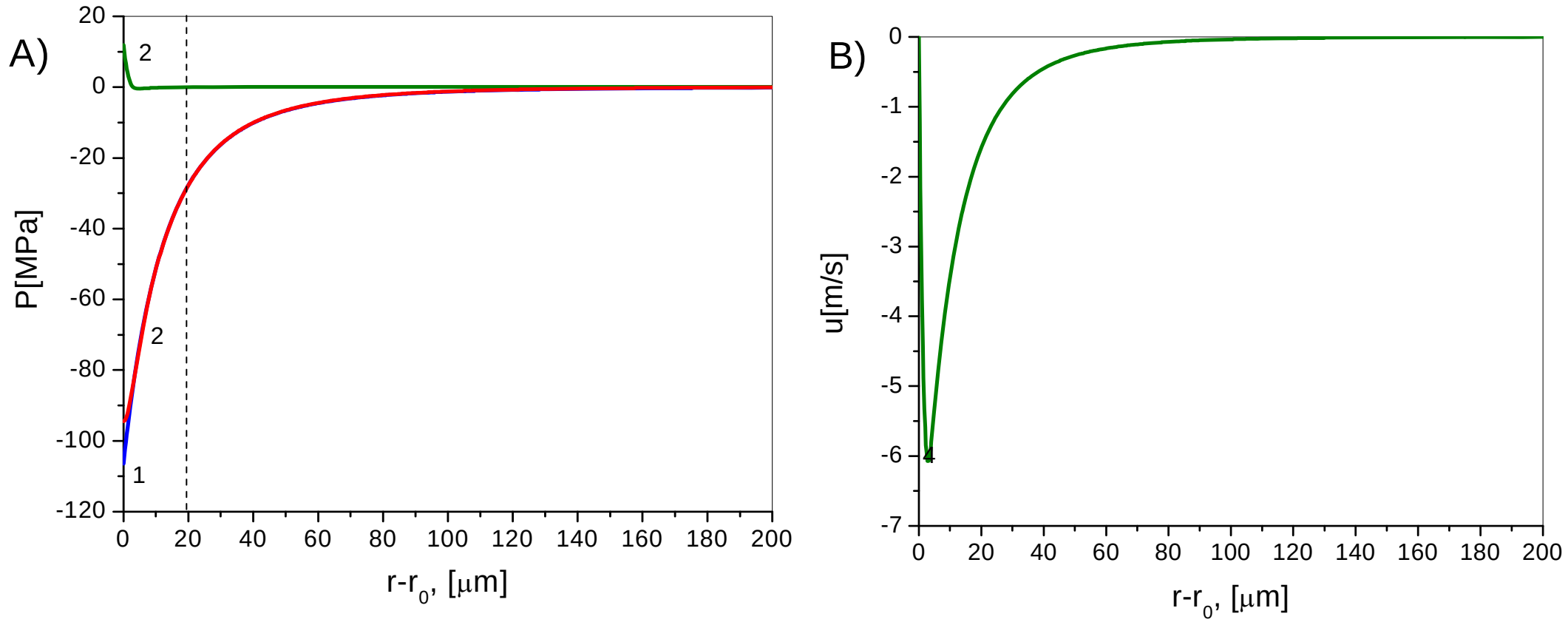

**Figure 5.** A) Pressure dependence on the distance from the electrode. 1 - is the pressure associated with the electrostriction forces, 2 - hydrostatic pressure, 3 - the total pressure equal to the sum of electrostrictive and hydrostatic pressures. The line indicates the distance, at which the pressure is greater than the critical, assumed, for example, $P_{cr}$ ≈ -30 MPa. B) Dependence of the velocity of fluid from the electrode distance at $t$=$t_0$=3ns.

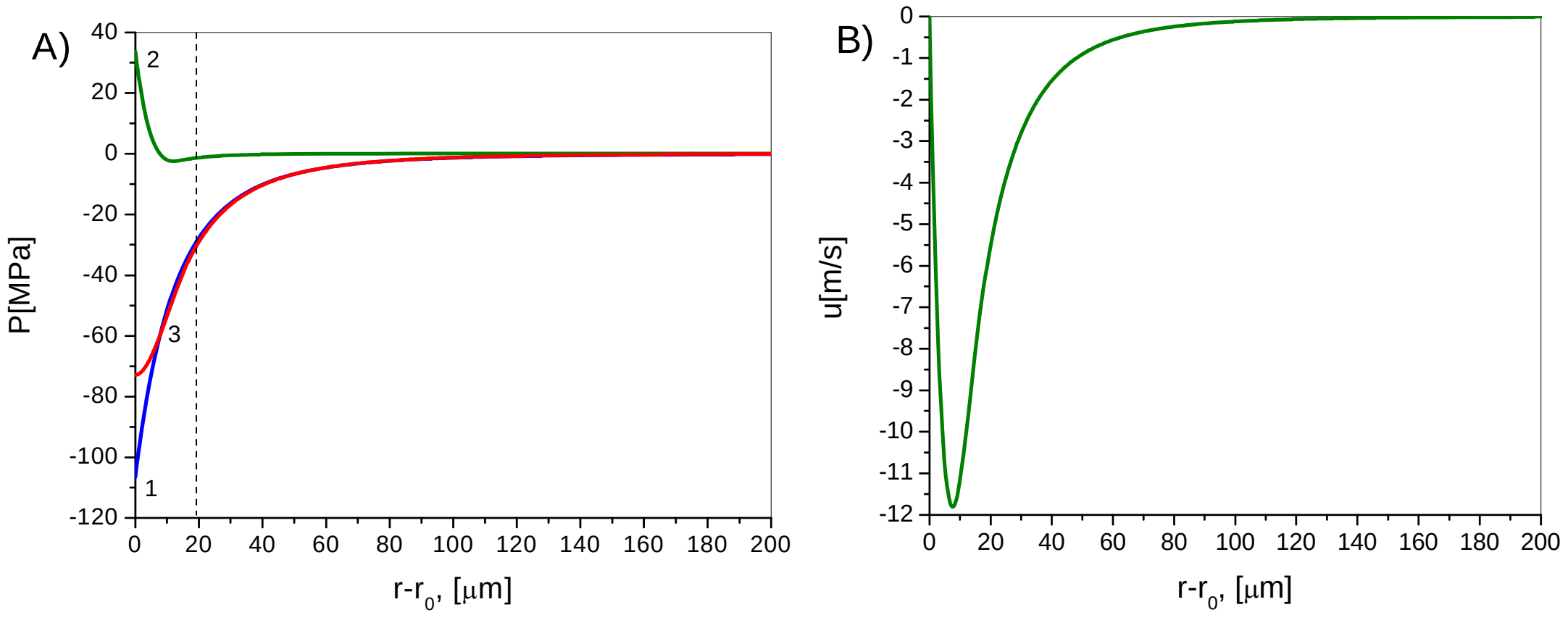

**Figure 6.** Same as in Fig. 5, at $t$=$t_0$=10ns.

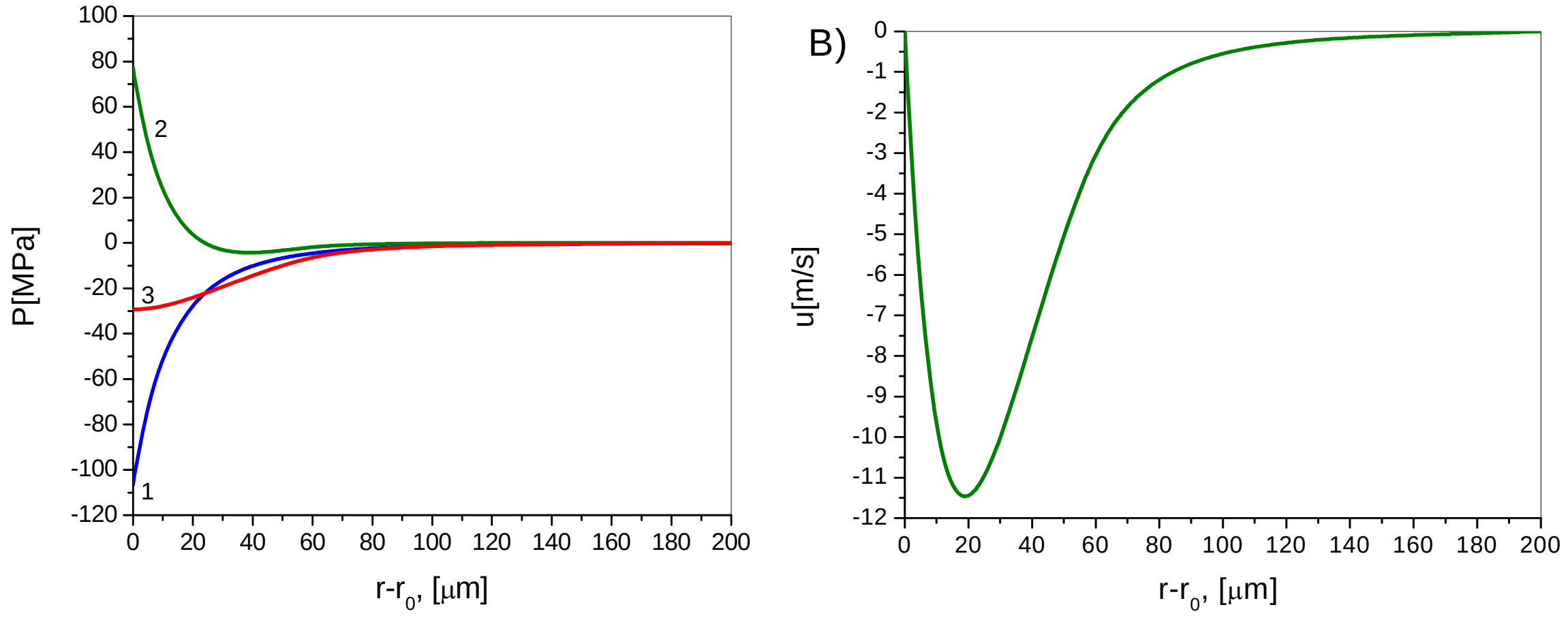

**Figure 7.** Same as in Fig. 5, at $t$=$t_0$=40ns. There is no region of developed cavitation.

For example, if the critical pressure, at which cavitation begins is $P_{cr} \approx -36.6$ MPa, obtained by comparing the optical measurements (see next section) with the results of calculation for the known values of the voltage amplitude, pulse duration, and electrode geometry, then $k_\sigma \approx 0.19$, $R_{cr} \approx 1.5$ nm, and the Tolman coefficient $\delta = 3.18$ nm. That allows calculating the probability of cavitation, the region where cavitation develops, and the concentration of the cavitation ruptures generated during the voltage pulse. At the same time, $\widetilde{P}_{cr} \approx (\alpha - 1)P_{cr} / \alpha \approx 0.33 P_{cr}$. The dependencies of the rate of generation of cavitation voids (12) on the distance from the electrode at different time moments during the voltage pulse at $t_0 = 3ns$, $r_0 = 50\,\mu$m, $U_0 = 22$kV, computed for $\widetilde{P} \geq \widetilde{P}_{cr}$, are shown in Figure 8.

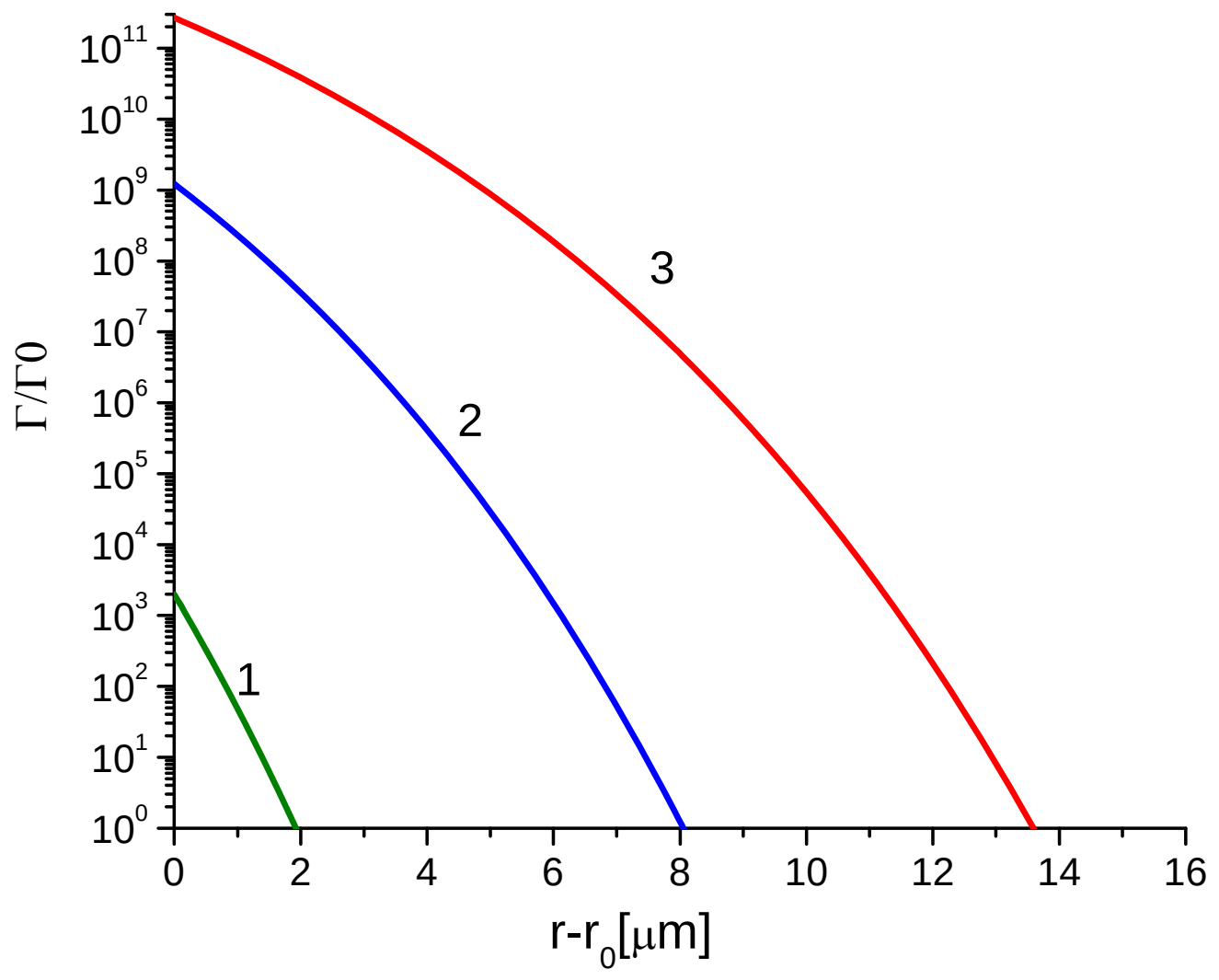

**Figure 8.** The computed dependencies of the rate of generation of cavitation voids $\Gamma(r,t)$, normalized by $\Gamma_{norm} = \Gamma(\text{at } P = -36.6\text{ MPa}) \approx 6.5 \cdot 10^{26}$ m$^{-3}$s$^{-1}$. The curve 1 corresponds to t = 2 ns, 2 to 2.5ns, 3 to 3ns

At a sufficiently steep voltage pulse, the fluid does not have time to shift to the electrode due to inertia. In this case, the negative pressure region substantially coincides with the stretching liquid. At longer voltage pulse, the front liquid has time to move into the region with a higher electric field. As a result, the area of negative pressure is very different from the case of "instant" turn of the electric field, and cavitation may not develop (as shown in Figure 7, which corresponds to the long voltage rise time, $t_0$=40ns).

It should be noted that we considered the case of spherical electrode only as an example of the numerical modeling, which allows for determining the area of negative pressure as a function of pulse amplitude and shape of the electrode. The conditions when the pulse is short enough so that the fluid does not have time to move to the electrode and compensate for the negative pressure associated with electrostriction have been considered in [14].

## 5. Experimental detection of the origin of the cavitation

A simple optical method for identifying areas of cavitation inception in a strong non-uniform electric field had been proposed in [15] on the basis of shadowgraph techniques [32]. Since the sharp boundaries of the cavitation bubbles (voids) lead to intensive scattering of light (some analog of opalescence [33]), scanning the laser beam in the vicinity of the needle electrode and observing the changes in the scattering characteristics can define the boundary of the region of cavitation. It is convenient to project the passed laser beam and the scattered radiation on a remote screen (Figure 9). By subtracting from the illumination on the screen (Figure 9, B) the initial unperturbed illumination (Fig. 9a) it is possible to locate a border of the cavitation region. By using the model similar to [14] the value of the negative pressure, at which the cavitation voids generation begins, can be found. Then, using the method of calculation of the probability of the cavitation described in section 1, we can determine the free parameters for the theory of cavitation voids formation. By reducing the voltage on the electrode, while performing the described optical measurements, the negative threshold pressure, at which cavitation voids appear, can be determined. The minimum voltage, at which the illuminance on the screen B) will not be different from the illumination on the screen A) allows finding the critical negative pressure, below which there is no cavitation.

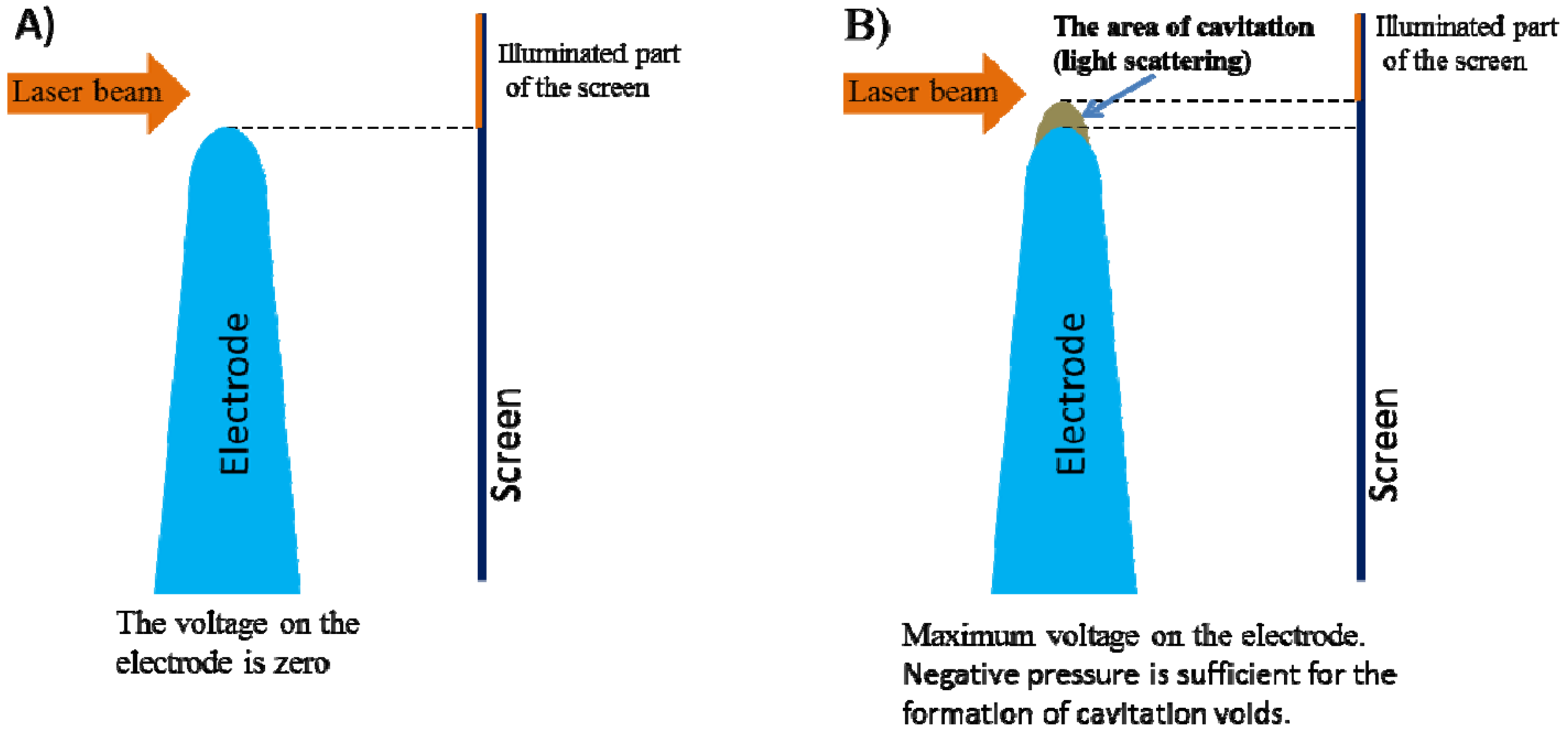

**Figure 9**. Optical observation of the cavitation region.

## Conclusions

At a quite steep voltage front (of the order of a few nanoseconds or less) applied to the needle electrode, the fluid does not have time to move. As a result, a significant region of negative pressure, in which the cavitation could develop, appears in the vicinity of the electrode.

The hydrodynamic model of dielectric fluid motion in a strong inhomogeneous non-stationary electric field, proposed in [14], allows to find the distribution of pressure in the fluid for any electrode and a voltage pulse, and to calculate the probability of the cavitation formation for any particular model of cavitation voids generation.

A comparison of the experimentally observed dimensions of the region where cavitation develops in the controlled unsteady conditions with the results of calculations in the framework of the hydrodynamic model [14], allows to obtain the critical parameters of cavitation initiation, which can be used in many problems of applied hydrodynamics.

## Acknowledgement

We are grateful to S.M. Korobeinikov for stimulating and fruitful discussions.